# VR-based Intervention for Perspective Change: A Case to Investigate Virtual Materiality

Ali Arya*    Anthony Scavarelli*    Dan Hawes*    Luciara Nardon**

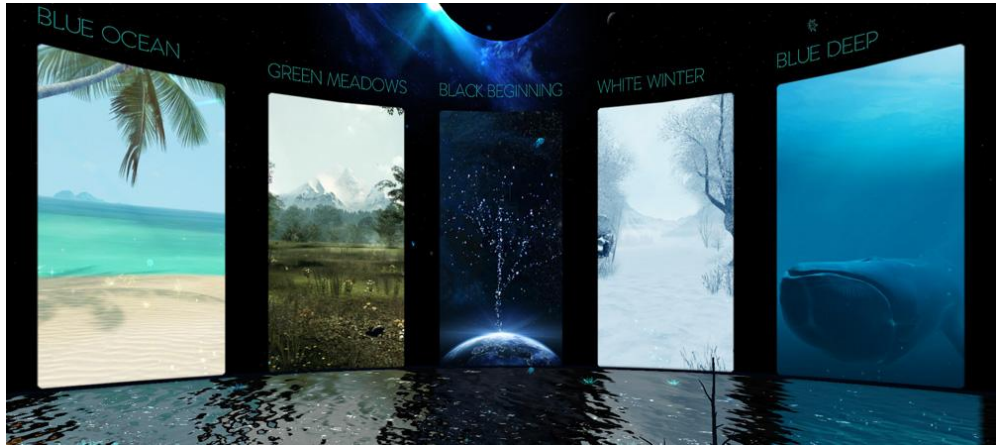

Figure 1: The menu screen from the Nature Treks VR app used by the participants

## ABSTRACT

This paper addresses the concept of materiality in virtual environments, which we define as being composed of objects that can influence user experience actively. Such virtual materiality is closely related to its physical counterpart, which is discussed in theoretical frameworks such as sociomateriality and actor-network theory. They define phenomena in terms of the entanglement of human and non-human elements. We report on an early investigation of virtual materiality within the context of reflection and perspective change in nature-based virtual environments. We considered the case of university students reflecting on the planning and managing of their theses and major projects. Inspired by nature's known positive cognitive and affective effects and repeated questioning processes, we established a virtual reflection intervention to demonstrate the environmental mechanisms and material characteristics relevant to virtual materiality. Our work is a preliminary step toward understanding virtual materiality and its implications for research and the design of virtual environments.

## CCS CONCEPTS

•Human-centered computing~Human computer interaction (HCI)~Empirical studies in HCI

•Human-centered computing~Human computer interaction (HCI)~Interaction paradigms~Virtual reality

*   School of Information Technology, Carleton University, Ottawa, Canada
** Sprott School of Business, Carleton University, Ottawa, Canada
arXiv Pre-print
Corresponding Author: Ali Arya, arya@carleton.ca

## KEYWORDS

virtual reality, virtual environment, reflection, perspective

## 1   Introduction

There is a large body of research that highlights the effect of the environment on cognition and emotion [5,33] and a growing number of studies that emphasize the active role of material objects and their effect on human experience [32,41], proposing the notion of sociomateriality [32]. Sociomateriality sees the material world's effects as significant as the social context. According to this research, the material world is not just a collection of objects that humans manipulate but a set of (somewhat) active agents that contribute to user experiences through entanglement and intra-action (a term that aims to highlight a bi-directional relationship that can go beyond human-centred manipulations and interactions) [32]. Materiality refers to the collective characteristics of environmental objects that give them a level of agency to initiate and participate in the intra-actions with human agents. For example, Hultin [20], when observing the interaction of newcomers with immigration officers, noticed how setting up a glass shield halfway through the research study significantly affected the newcomers' experience without having been considered in initial research assumptions.

Popularized by many researchers from psychology [13] to design [31] and human-computer interaction (HCI) [12], an affordance is a property of an object or environment that is "compatible and relevant to people's interaction" [12]. Virtual reality (VR) has a unique combination of affordances such as immersion, visualization, and interaction [37] [39], which have been shown to make VR an effective tool in applications such as learning [15], well-being [40], and increasing awareness and empathy [17]. In this paper, we address virtual materiality as a less-investigated affordance of virtual environments (VEs), defined as being composed of objects able to influence user experience in an active way, closely related to its physical counterpart. Our notion of virtual materiality differs from those based on the common usage of the word material in 3D graphics, referring to the visual surface properties of a 3D model [4]. Also, we use the terms VR, VE, and 3D VE interchangeably, even though they may have different meanings in other contexts.

There is considerable research on the cognitive and emotional effects of VEs on users [8,10,11,15,16,17,18,19]. However, most of them investigated environmental factors and elements that were specifically designed to influence users and their performance of specific tasks, such as learning a topic, playing a game, or exploring space. While entanglement with material (and other non-human) elements is suggested as the basis for the fourth wave of HCI research and design [9], investigating the effect of virtual environments on users' cognition, emotion, and experience has been mostly limited to objects and settings that are controlled and planned for a specific effect. This limitation does not properly account for and allow the study of materiality in virtual environments and how it can initiate different, unplanned, and unexpected experiences, as the environmental parameters are highly controlled, and observations are for pre-defined metrics.

We propose the notion of *virtual materiality* defined as an affordance of VR that offers a virtual counterpart to the intra-active material objects in physical environments. Such an affordance suggests that a VE can be considered as a complex set of agents affecting the user experience beyond the initial intent and in unexpected manners. For example, a virtual room may have been created for a virtual classroom application, but the design may have initiated certain cognitive or affective processes in users that were not expected or planned [16]. While materiality in physical environments has been investigated and examples of it have been mentioned in virtual environments, there is not sufficient knowledge on whether or not virtual materiality acts in a way similar to its physical counterpart. The unique characteristics of VR/VE suggest that the mechanisms for virtual materiality may be different from the physical one or have unique and different features, which warrants specific and dedicated research. The study of virtual materiality requires extensive research on the effects of existing and newly designed VEs on users to understand how VE objects could initiate unplanned influences on users performing different tasks. While such a characteristic can complicate the design of VEs, it can offer new possibilities, as do the physical counterparts in the physical world vs. controlled lab environments.

We briefly discuss potential research and design implications of virtual materiality in this paper, after presenting our research study, as a preliminary step towards investigating materiality in virtual environments. Our study was focused on a specific set of VEs and user tasks to be performed in them. We chose reflection and perspective change, through an established repeated questioning intervention [44], for university students working on their theses or major projects, as an initial context of our investigation. This choice was motivated by various cognitive and affective processes that could be affected by the environment and the existing literature providing a theoretical grounding for such effects [1,3,5,6]. We also chose different nature VEs as examples to use due to research suggesting generic natural settings can act as catalysts for reflection and overall well-being [6,11,29]. By generic, we mean general purpose, not planned for the specific task. Examples can be using a common environment (e.g., a park or other general-purpose setting) for a particular activity (e.g., learning) or being exposed to dynamic, unpredictable, and unintended situations (e.g., seeing a flock of birds). Our primary research question is how (through what mechanisms and material characteristics) VEs affect participants' perspectives and approaches on a topic/problem they are thinking about (or are struggling with) and how they approach and solve it. As a secondary question, we also wanted to see if VE-based intervention has the reflective effect of traditional methods and allows users to go through the expected stages. More explicitly, exploring the reflective effect will enable us to verify that the environmental effects were, in fact, positive and not a distraction in the process. Due to the generic nature of the environments, we hope that our findings, as initial as they may be, can help understand how materiality works in VEs and how it can be considered an additional affordance.

## 2  Related Work

Many VR researchers have directly investigated the effect of virtual environments on thinking and learning. For example, Rizzo et al. [34] discuss the effect of VR in clinical situations and how it can be used for cognitive rehabilitation, and other researchers have investigated using VR as an experiential learning tool [7,36]. Most of these examples focus on environments intentionally designed to assist with primary tasks such as learning [7]. Research on the effect that random and generic parts of a VE can have on users is mostly focused on natural settings, owing to the body of research related to the positive effect of nature on well-being [1,6] and thinking/cognition [3]. Research suggests that virtual environments can have spiritual and emotional effects similar to those of their physical counterparts [30]. VR researchers have investigated the effects of virtual nature exposure and showed the potential positive role such exposure can play in emotional well-being [8,10,21,25,29,38]. Despite these examples, the effect of generic environments (such as nature) on thinking and perspective change is not adequately investigated in VR. Investigation of virtual materiality should go beyond overall well-being and emotions and consider more specific aspects of the experience, such as cognition and behaviour, and the mechanisms through which the VE affects them.

While theories such as situated cognition do discuss the dependency of cognition on context, alternative perspectives, such as sociomateriality [2,32], go beyond dependence and suggest entanglement., i.e., a close inter-dependence between human(s) (i.e., social) and objects including nature, technology, etc. (i.e., material). Such interdependence requires the elements to be studied together as a phenomenon rather than in isolation. Entanglement and related theories, such as Actor-Network Theory (ATN) [23], are considered as the basis of a fourth wave for HCI [9]. Sociomateriality offers a new ontological perspective that gives material objects a key agential ability, and as such is a suitable framework for studying material characteristics and effects of VEs. Karen Barad [2] introduced agential realism, the foundation of sociomateriality, as a relational or "becoming" ontology, which posits that "the world is an ongoing open process of mattering" [2]. Being relational means that the primary unit of ontology is not the object but the relations. Objects' identity and properties are dynamic and are defined through their relationships. Becoming or mattering happens through intra-actions, a term that implies that actors are not independent but defined within entangled relationships, affecting each other. The agency here does not mean that, for example, a book writes itself, but that all actors in a writing process influence each other; the writer is influenced by the technology they use and the surrounding objects as much as they change them [2].

## 3 Study Design

Our study investigates the effect of the virtual environment (Figure 2) on reflective thinking in the context of university students dealing with challenges in planning and managing their theses or major projects. We used a repeated questioning process called Clean Networks [43,44]. Repetitive processes are shown to help with reflection and perspective change [22,28,35]. Clean Networks is an intervention based on Clean Language, a coaching approach designed to guide a coach to communicate with a client [42,44]. The key concept in this practice is to avoid any input from the coach that can influence the client's thinking. So, it relies on "clean" questions and comments, i.e., those causing the least bias (a.k.a. contamination). The Clean Networks method is particularly suitable for the study of materiality as it uses environmental effects by repeatedly asking the participants what they know about the topic of reflection while moving to a new location/perspective after each answer.

### 3.1 Participants

Our study was approved by the institutional research ethics board. We invited students at the research team's university to participate and ran the study with 21 participants with an average age of 27.8 (ranging from 18 to 44). Eight identified as man and thirteen as woman. The numbers of undergraduate, Master's, and PhD students were 5, 7, and 9, respectively, coming from different fields of study. Two participants considered themselves experienced with VR, while 9 had limited familiarity and 10 had no experience with it. Similarly, 4 were experienced in formal reflection, 9 had some and 8 had no experience with it. Our findings showed no significant relation between age, gender, and VR/reflection experience and the results discussed in Section 4.

However, our number of participants is not large enough for such statistical analysis.

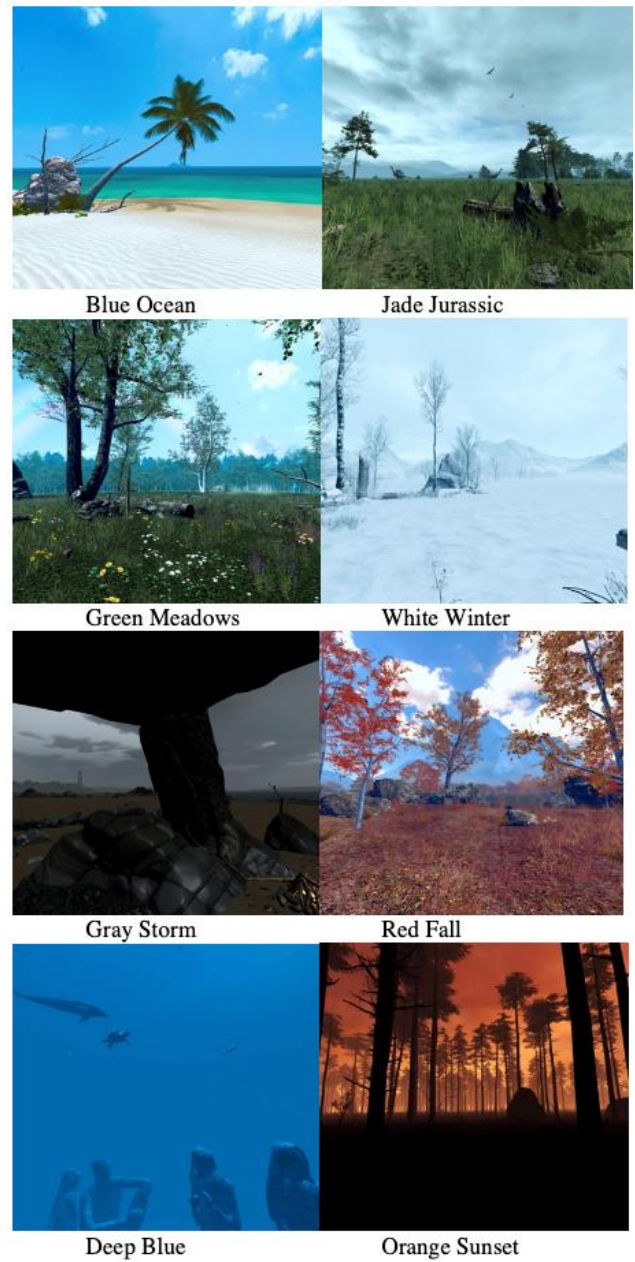

**Figure 2. Examples of Virtual Environments**

### 3.2 Apparatus

The study was performed in person using the Meta Quest 3 VR headset. We used the Nature Treks VR app, available on the Meta Quest Store (Figure 1). Participants could choose to visit a variety of virtual environments as illustrated in Figure 2. Since the primary data collection was through observation and interview, no survey, questionnaire, or other instrument was used. The conversations with the participants were audio recorded using a mobile phone or laptop, and no other instrument was used. The participants used Quest controls for interaction. On the right hand, they only used the trigger button to select a virtual environment and navigate using teleportation. On the left hand, they used the menu button to return to the main menu. No other buttons and interaction methods were used. The environments were very interactive in the sense that they provided plenty of opportunities to navigate, view, get closer, and occasionally pick up objects in a wide range of situations. As initial research, we wanted to focus on a limited and basic set of interactions, not including manipulation of the environment or influencing the elements. Such interactions can be included in future research.

### 3.3 Procedure

We designed and followed a VR version of the Clean Networks process. Participants attended the session in a lab on the university campus, with the following steps taking about 30-45 minutes:

1. Briefly explaining (to the researcher) the academic goal they would like to reflect on.
2. Using the VR HMD, briefly exploring the virtual environments to get comfortable with the procedure.
3. Going to the main menu, thinking about their reflection topic, and choosing an environment option.
4. Navigating within the environment, choosing and stopping at a location. Then, answering two clean questions with proper pauses between them for looking around, getting comfortable, and thinking:
   i. What do you know about [topic] from that … space … there?
   ii. Is there anything else?
5. Moving to a new location by following the instructions:
   i. Find another place in this or in a different environment that feels right
6. Repeating steps 4 and 5 above until six rounds of questions were completed.
7. After 6 rounds, going back to the main menu and answering the final concluding clean questions:
   i. And what do you know NOW about [topic]?"
   ii. And is there anything else you know… now?
   iii. And what difference does it make?
8. Removing the HMD and answering a few short questions about how they feel and think about this experience.

### 3.4 Data Analysis Method

We used a mixed method (mostly qualitative with some quantitative data collection and analysis) approach where we collected data through interviews (clean questions and background information) and the observation of the virtual environments the participants visited and their emotional reactions. As the work reported here was an initial step, we analyzed the data through a top-down deductive coding for concepts already identified in existing literature [27]. These were related to stages of the reflective process and environmental influences on thinking, with the goal of establishing associations between questions, environment, and the participants' cognitive process (i.e., how their ideas on the subject evolved). As discussed later, after our analysis, we realized that further coding and analysis is necessary with more focus on material characteristics of the environment, which will be the topic of future work.

Based on the commonly mentioned mechanisms through which the environment affects thinking [5,33], we considered the following environmental effects: Association, Triggered emotion, Recalled memory, Metaphorical thinking, Ethical/moral consideration, Social relations, and Embodied cues. These formed the set of codes for the first dimension of our study, *World Analysis*, where world refers to the virtual environment for each stop. These codes were to help answer our primary research question on how the VE influenced the reflective process. In coding for world analysis, we paid attention to actual environmental characteristics causing these effects, such as the shape, colour, and nature of objects.

We considered different models of reflective thinking used in Clean Language coaching [14,24] and drew on these models, and chose five key aspects for the reflection process: Problem (what the participant doesn't like), Outcome (what the participant wants to happen), Resource (anything that can help the participant reach the desired outcome), Change (any explicit change in thoughts and feelings while going through questioning), and Action (anything the participant decides to do). Based on our pilot tests, we also considered Comment (general descriptions of the VE or participant's background, not directly related to the above stages) and Wobble (inability to say or think of anything) [14]. These formed the set of codes for the second dimension of our study, *Round Analysis*, where round refers to each stop at VE by a participant. This analysis was to answer our second research question to see if the VE helped follow the expected reflective process, as in traditional Clean Network approach.

## 4 Results and Analysis

The interventions were done by three members of the research team, who each did the first iteration of coding for their corresponding data. We analyzed two samples together first to make sure we all had the same understanding of the codes and made some minor adjustments. The fourth member of the team then did a second iteration of coding, and we compared the results, discussed to make sure no evidence of code occurrence was overlooked or mistaken and reached a conclusion on all items.

As common and accepted in HCI qualitative research [26], we did not calculate and emphasize quantitative inter-coder reliability measures, as we wanted to encourage different insights and discuss them. The primary sources of initial difference were inherent vagueness in the nature of data and the possibility of more than one code being present in any round where we were to choose only one as the dominant. For example, the code Resource was supposed to be used when the participant identifies anything that can help them achieve the desired outcome, including people, ideas, tools, and feelings. The following text was considered by one of the coders are a Resource:

> "I can observe other kinds of life and other ways to live, another time to leave, another frequency, and it's a good feeling."

But others pointed out that while it shows a positive and potentially helpful emotion, it is not directly pointed at solving the problem. It is more about how the participant feels about the environment. So, the code was assigned as Comment as it doesn't directly map to any of the major codes (Problem, Outcome, Resource, Change, and Action).

### 4.1 Summary of Coding

From the data, we observed that the reflective process in VR is working as expected, i.e., earlier rounds include a higher occurrence of Problem and Comment, and moving forward, we could see more Outcome, Resource, and then Change and Action. Most participants (18 out of 21) concluded with Change or Action in round 7. The data suggests that the process was somewhat effective in reducing the relative number of Comments in favour of identifying Problems, Outcomes, and Resources. Overall, round analysis showed a process similar to what was expected in non-VE cases. This answered our secondary question, which is not the focus of this paper and acted mostly as check point. The specific and potential role of VR in reflective process will be the topic of another research path.

We considered Association as a comprehensive term that shows establishing any connection between the elements of the virtual environment and any concept related to the topic of reflection. As such, it included other codes such as memory, emotion, and metaphor. This resulted in the number of association codes being much higher than the others, as expected. While we saw world codes happening with all of the round codes, some specific observations can be made from Table 1.

Table 1: Round vs. World Analysis Codes. The number of world analysis codes occurring with each of the round stages

| Round | Association | Memory | Emotion | Metaphor | Moral | Social | Embodied | Total |
|---|---|---|---|---|---|---|---|---|
| Problem | 8 | 4 | 3 | 6 | 1 | 1 | 2 | 25 |
| Comment | 16 | 11 | 11 | 13 | 0 | 3 | 8 | 62 |
| Outcome | 7 | 4 | 5 | 7 | 2 | 4 | 1 | 30 |
| Resource | 13 | 4 | 9 | 11 | 1 | 8 | 2 | 48 |
| Change | 18 | 11 | 11 | 14 | 0 | 8 | 6 | 68 |
| Action | 11 | 4 | 11 | 10 | 1 | 2 | 1 | 40 |
| Total | 73 | 38 | 50 | 61 | 5 | 26 | 20 | 273 |

Metaphors were clearly the highest occurring influence mechanism, followed by emotional responses and memories. Problem and Outcome co-occurred with fewer world codes, which makes sense as the participants did not need to dig deeper or be influenced by the environment to state their challenges and desired situations. Resource, Change, and Action, on the other hand, happened together with many more VE-triggered phenomena, which can suggest the VE was influential in reaching changes or identifying actions and resources as intended. Moral, social, and embodied triggers happened less frequently, in general, than other world codes. This did not surprise us as the VEs were single user, with no human element, and had limited interactions (mostly walking and looking around). Social concepts were triggered through metaphors initiated by animals or environmental objects. Embodied effects were experienced through actions such as walking, going to higher elevation, being surrounded by water, and sitting down. Overall, we observed that VEs with more active elements, such as Blue Deep and Jade Jurassic, had more occurrences of world codes. The coding process illustrated how (1) the participants' thought process was progressing through the stages defined in the round analysis and (2) certain characteristics of VEs were influencing them through mechanisms in world analysis code, as illustrated in the next section.

### 4.2 Environmental Effects

The objects in VE influenced the user experience and reflection process through their material characteristics and the mechanisms identified in our world analysis. Following, we illustrate these influences grouped by those mechanisms. In the examples, the material characteristics that initiated the effects are **boldfaced**.

**Association**

While some participants expressed their challenges without any association with the VE, some benefitted from such associations to clarify, for themselves and us, what is troubling them. In the following two examples, associating with the **arrangement** or **size** of objects helped participants understand and express their current situation and problem:

> Q: And what do you know, from that space there?
> A: In this space, I know that I'm kind of stuck between two rocks. That one, maybe, represents my kind of vision of *what I want my research to be. And the other rock sort of represents what's realistic, given the time and scope. So, it's kind of like I'm stuck here having to figure out how to get unstuck while also maintaining my, I guess, core values.*

Q: And what do you know about your thesis from there?

A: *Looking at the ocean, I think, maybe, tells me about how big of a task it is.*

**Memory**

Many cases of association specifically involved recall of past memories that were effective in various stages of reflection. For example, in the following case, **natural scenery** reminded the participants of their roots and motivations, and so helped with a change of perspective and identifying new emotional resources.

*So, I think this place reminds me a little bit of Canada, obviously with the snow, and the mountains and the animals. But the thing that came to my head when I entered this area was Why, why am I doing the PhD in the first place? I think one of the main reasons I've started it was to kind of raise the expectations for my sons. Show them that they are capable of doing something difficult, hard, something intellectually challenging.*

**Emotion**

Similar to the previous example, scenery and especially **lighting** caused emotional responses, which in turn helped participants identify desired outcomes, e.g., "joy in research" in the following case:

*Oh, well, that first one, this world looked really pretty. But when I popped in, I was like, Oh, this is kind of scary. But now that I'm here, I can see like, the lake is really pretty. And these kind of like Northern Lights, or something, are really pretty and it's kind of the music is nice kind of, actually nice and serene. Which I think is making me remember when I first started [my research], how like nervous I was and everything. And I think it's also important to remember the joy in research. [laughter] And kind of allow, maybe allow myself to be excited, instead of so focused on deadlines, and employability and all these things around the thesis.*

The following participant experienced a **relaxing** and **calm** virtual environment that helped them discover the importance of setting up a happy and relaxing workplace as an action for achieving their goals:

*I feel like this is the optimal place. You could do things in different ways, like my thesis, the quality will be different. This one is the optimum one where all the conditions are met then you could end up not only finishing the thesis but finishing with a good quality. You need to be happy, I think, if I'm to make progress. That's the idea that's getting to my head now. So, you really have to maybe look at what makes you happy and develop environment that makes you happy and then you start making progress.*

**Metaphor**

A common and particular example of making associations was through metaphors. **Actions, arrangement**, and **status** of VE objects provided opportunities for metaphors which helped identify resources in the following cases:

*These two dinosaurs are kind of far from that one. [laughter] And they're together and they're kind of not caring. And they're wandering together and it kind of makes me think like I guess, to focus on my thesis and graduate on time. You know, I just have to keep going. Even if you know, the threat of not graduating on time, or things like, gives me anxiety, I can, you know, I still have to keep living and surviving and moving. And there's other people like [X] and [Y] that are there to support me through it as well and to help me and that are like on this journey with me.*

*All these tall vertical trees, remind me of researchers in this area and each area to some extent that are kind of like pillars of that particular research stream. So, if I'm struggling to find relevant articles, a useful thing to do is always go back to their papers and look at the people that they've referenced. And also recognizing to some extent of the people on my committee fit that mold. So, I can always reach out to them if I have questions or concerns.*

**Moral/Value**

Particular **settings** and **arrangements** caused participants to consider moral issues and personal values and so changed their perspectives on the challenge they had, as in the following example:

*I feel like a concert because I'm on the stage or podium. It's weird because at the same time, nobody's here. Just my journey. And it's interesting because I can remember that my journey is about me, not about another person or people and I need to put my focus on myself.*

**Social**

**Interacting elements** or even **remains of human manipulation** initiated social considerations resulting in different forms of change in perspective or identifying resources.

*This [fire pit] gave me the evidence that another person was here. It's making me think what I'm forgetting is that not only there are people around me while I go through this, but there's other people who have been in my position who have gone through this and who now have great careers. And it's a nice reminder.*

**Embodied**

Many aspects of the VEs facilitated embodied experiences through a sense of **motion**, **elevation**, **immersion**, etc. Through these experiences, the participants felt more connected to the environment and were able to comment on it and then on their particular challenges.

*Let's find out. Can I get in this boat? No. I gotta look up. Can I climb? Oh, my gosh, I can climb. Oh my god. [Excitement] [Scream] I'm on the floor. I'm not gonna fall. Oh. [expressing excitement] This is too crazy. Do you feel nervous? I am sweating! I realized that maybe I do have a fear of heights. Crazy!*

*I really like being by the water. I like to swim. I think it's a good place to read also it looks pretty. I like how the water is*

*always moving so I don't know makes me think about life how nothing is static everything is moving.*

## 5 Discussion

In response to our primary research question on the role of materiality in virtual environments, we demonstrated that different elements in VE could trigger and initiate thoughts and emotions in the user, even in cases where these elements were not specifically planned for that purpose. This initiation happened through mechanisms such as memory, metaphor, and embodiment, and was related to material characteristics such as size, shape, arrangement, lighting, and scenery. The affected thought processes (in our case, reflection) could lead the user through a discovery of novel insights (i.e., gaining new perspectives) on the topics they found challenging. Our findings are twofold. First, they show the ability of generic VEs to act as a catalyst for reflection and perspective change. Second, they emphasize the role of such material objects and settings in the overall virtual experience.

Based on our findings, we suggest further investigating virtual materiality as a counterpart to, or extension of, sociomateriality in VEs. Such virtual materiality may posit that despite their limitations in terms of interaction and modality, VEs are made of objects (material entities) that are not merely acted upon. They can actively shape users' experience based on different characteristics and through different mechanisms without being intended for that effect. A virtual phenomenon, in this sense, can be an intra-action (in a sociomateriality sense) between entangled entities, just like its counterpart in the physical world.

Knowing that random virtual environments, not designed for any specific purpose, can trigger new thoughts means that there is a new way of helping individuals with their reflective goals. This is particularly important once we consider the ability of automated systems to implement "clean" repeated questions. This idea can then be extended to other cognitive tasks, especially learning, where a learner or entire groups can practice their activities within different environments. A balance between positive influences as observed in our study and negative ones such as unwanted distraction should be the subject of further research, so should the suitability of different environments for different tasks.

On the other hand, the materiality of virtual environments and the effect of VEs on user's mood and cognition can raise significant concerns about the way VR design and research is conducted. If a small object can trigger significant cognitive and affective consequences, the validity of many research findings and design decisions based on controlled settings will be questionable or limited. Again, further research is needed to investigate the way materiality presents itself in virtual environments and how it should be considered in future work.

Our analysis was preliminary, as we only coded for the environmental influence mechanisms mentioned in the literature, in a purely deductive way. Further deductive and inductive (i.e., abductive) research can look for other mechanisms and also code for specific elements. Also, due to logistic reasons and the preliminary nature of this study, we only investigated natural settings. Urban, abstract, and other forms of environments can have their own materiality and influences, which were not considered in our study. Similarly, other tasks and experiences beyond academic reflection, and other forms of interaction with the environment, should be included in investigating materiality. It will also be helpful to have a non-virtual baseline for comparing how effective VR is for reflection, but our focus on this preliminary research was only to demonstrate that material characteristics of VEs could meaningfully change perspective and influence user experience.

We further recognize that the limited number of participants does not allow generalizations and that the occurrence of the world effects at the same time as reflective stages does not mean direct causation. However, we believe our qualitative data suggests a link and is a motivation and useful starting point for future research.

## 6 Conclusion

Our study was an exploratory investigation of how VR can include materiality as a significant characteristic and how such virtual materiality can influence user experience in unexpected ways. We illustrated this through the ability of generic environments to guide participants through a repeated questioning process and allow them to reach new insights through exposure to random virtual stimuli. Based on the findings, we introduced the notion of virtual materiality and reviewed some possible implications it may have. Future research is needed to overcome the limitations of our study and offer more insights into the role of materiality in virtual environments.